\documentclass[english,aps,prl,twocolumn]{revtex4-1}
\usepackage[T1]{fontenc}
\usepackage[latin9]{inputenc}
\usepackage{float}
\usepackage{graphicx}
\usepackage{esint}

\makeatletter
\@ifundefined{textcolor}{}
{%
 \definecolor{BLACK}{gray}{0}
 \definecolor{WHITE}{gray}{1}
 \definecolor{RED}{rgb}{1,0,0}
 \definecolor{GREEN}{rgb}{0,1,0}
 \definecolor{BLUE}{rgb}{0,0,1}
 \definecolor{CYAN}{cmyk}{1,0,0,0}
 \definecolor{MAGENTA}{cmyk}{0,1,0,0}
 \definecolor{YELLOW}{cmyk}{0,0,1,0}
 }

\makeatother

\usepackage{babel}
\begin{document}
\global\long\def\bra#1{\left\langle #1\right|}

\global\long\def\ket#1{\left|#1\right\rangle }

\global\long\def\bk#1#2#3{\bra{#1}#2\ket{#3}}

\global\long\def\ora#1{\overrightarrow{#1}}

\title{Signature of Phase Transitions in the Disordered Quantum Spin Hall State From the Entanglement Spectrum}

\author{Matthew J. Gilbert$^{1,2}$,  B. Andrei Bernevig$^{3}$, and Taylor L. Hughes$^{4}$}

\affiliation{$^1$Department of Electrical and Computer Engineering, University of Illinois, 1406 West Green St, Urbana IL 61801}
\affiliation{$^2$Micro and Nanotechnology Laboratory, University of Illinois, 208 N. Wright St, Urbana IL 61801}
\affiliation{$^3$Department of Physics, Princeton University, Princeton NJ 08544}
\affiliation{$^4$Department of Physics, University of Illinois, 1110 West Green St, Urbana IL 61801}
\date{\today}

\begin{abstract}
Of the available classes of insulators which have been shown to contain topologically non-trivial properties one of the most important is class AII, which contains systems that possess time-reversal symmetry $T$ with $T^2=-1.$ This class  has been the subject of significant attention as it encompasses non-trivial Z$_2$ topological insulators such as the quantum spin Hall (QSH) state and the 3D strong topological insulator. One of the defining properties of this system is the robustness of the state under the addition of disorder that preserves $T.$ In this letter, we explore the phase diagram of the disordered QSH state as a function of disorder strength and chemical potential by examining the entanglement spectrum for disordered class AII symplectic systems. As for the case of the $T$ breaking Chern insulator we show that there is a correspondence between the level-spacing statistics of the Hamiltonian and that of the level spacing statistics of the entanglement spectrum.  We observe a feature in the statistics of the entanglement spectrum that aids the identification of delocalized states and consequently critical energies across which phase transitions occur.
\end{abstract}

\maketitle

The classification of topological insulators and superconductors via the periodic table\cite{qi2008b,Schnyder:2008gf,Kitaev} has garnered much interest over the last few years. One of the most exciting results to have come out of the classification is the connection between the ten symmetry classes of gapped free-fermion systems and the ten random matrix ensembles studied by Altland and Zirnbauer\cite{altland1997,Schnyder:2008gf}. The idea that each symmetry class of topological states has a corresponding random-matrix ensemble enables one to carefully study the properties of disordered topological insulators, and their corresponding localization/delocalization transitions between topological and trivial insulators. As an example, the phase diagram of the disordered Chern insulator\cite{haldane1988}, which belongs to the Wigner-Dyson unitary ensemble denoted by class A, has been calculated in Ref. \onlinecite{prodan2010}. The other experimentally relevant class is AII which is the symplectic class which contains insulators with time-reversal symmetry $T$ such that $T^2=-1.$ This class is known to support both 2D\cite{Kane:2005sf,bernevig2006a,bernevig2006c} and 3D\cite{hasan2010} topological insulators classified by a $Z_2$ topological invariant. 

Our focus will be on the 2D insulator of class AII which is also known as the quantum spin Hall  (QSH) state. The properties of the disordered QSH state have been previously studied in Ref. \cite{prodan2011} by numerically evaluating the Z$_2$ topological invariant while varying disorder strength, chemical potential, and the nominal insulating gap. We seek instead to study this system via the entanglement spectrum\cite{li2008,fidkowski2010,turner2010,prodan2010} in the hope of developing new tools to study phase transitions in disordered systems by only considering ground state properties and not the excitation spectrum.  Recently it has been shown that many properties of the entanglement spectrum mimicked properties of the real energy spectrum; specifically the statistics of nearest-neighbor level spacings\cite{prodan2010}. There are two additional parameters in the calculation of the entanglement spectrum with respect to the Hamiltonian spectrum: (i) the choice of the ground-state (single-particle energy level filling) (ii) the choice of the Hilbert space partition/cut. In this Letter we will show that, in the context of topological insulators, studying how the statistics of the \emph{entanglement} spectrum varies when the single-particle filling varies can reveal regions of delocalized states in the \emph{energy} spectrum in the presence of quenched disorder.

We will consider an effective square-lattice model for the 2D QSH Hamiltonian\cite{bernevig2006c}  as a representative Hamiltonian for the AII symmetry class. The Fourier transformed clean Hamiltonian is simply 
\begin{equation}
\label{eq:basicham}
H_{qsh}= \sum_{k}c^{\dagger}_k\left( \begin{array}{cc} H(k) & 0 \\ 0 & H^*(-k) \end{array} \right)c_k.
\end{equation}
with 
\begin{equation}
\label{eq:hdef}
H(k)= \vec{d}(k) \cdot \vec{\tau}
\end{equation}
where $c_{k}=(c_{A\uparrow}\;\; c_{B\uparrow}\;\;c_{A\downarrow}\;\; c_{B\downarrow})^T$ with $A$ and $B$ representing the orbital degree of freedom, $\vec{\tau} = (\tau^x,\tau^y,\tau^z)$ are the Pauli matrices for the orbitals, and $\vec{d}(k) = (\sin k_x,\; \sin k_y,\; 2-m-\cos k_x-\cos k_y).$  The topological state of interest occurs when $0<m<2,$ and, for example, a trivial state occurs when $m<0.$ A model similar to this was used as an effective model for HgTe/CdTe quantum wells\cite{bernevig2006c}. While this model represents a time-reversal invariant topological insulator, its random matrix properties when time-reversal invariant disorder is added \emph{exactly} match that of two copies of the Chern insulator and thus falls into class A. To study the properties of the symplectic class AII we need to deform this model, which we will do by adding spin-mixing terms which preserve T, notably a term which is due to the bulk inversion asymmetry (BIA) of the HgTe/CdTe zinc-blende lattice structure. We will choose the BIA term 
\begin{equation}
\label{eq:hbia}
H_{BIA}=
\left( \begin{array}{cccc}
0 &  0 & 0 & -\lambda \\ 0 & 0 & \lambda & 0 \\ 0 & \lambda & 0 & 0 \\ -\lambda & 0 & 0 & 0 \end{array} \right),
\end{equation}
where $\lambda$ is the strength of the BIA which was derived in Ref. \onlinecite{Taylortheory}. This term mixes spin-up and spin-down, preserves T, but breaks inversion symmetry; in turn it deforms our model away from a pathological case so that it represents class AII. The last piece of the Hamiltonian is the disorder which we take to be on-site potential disorder with a random strength uniformly distributed over $[-W/2,W/2]$ where $W$ is a real parameter.

With the Hamiltonian established, we now examine the Hamiltonian energy level statistics and the resultant entanglement spectra statistics. In collecting statistical data, we use a lattice model with size $40\times 40$ with periodic boundary conditions imposed on both the $\hat{x}$ and $\hat{y}$ directions so that there are no low-energy edge states and the system is completely gapped (in the clean limit). We use exact diagonalization to calculate the energy spectra and the entanglement spectra via Peschel's method\cite{peshcel}; essentially calculating the two point correlation function
\begin{equation}
\label{eq:corr}
C_{ij}  = \langle c^{\dagger}_i c_j  \rangle
\end{equation}
as a function of the lattice positions $i,j$ where the brackets $\langle \;\; \rangle$ mean the ground-state expectation value for a single disorder configuration. If one restricts $i,j$ to lie in a subset $A$ of the entire lattice then one can extract the single-particle entanglement spectrum by diagonalizing this truncated matrix. For our calculations we took the region $A$ to be a rectangular subset of size $20\times 20.$

\begin{figure}[t]
\centering{}\includegraphics[width=0.5\textwidth]{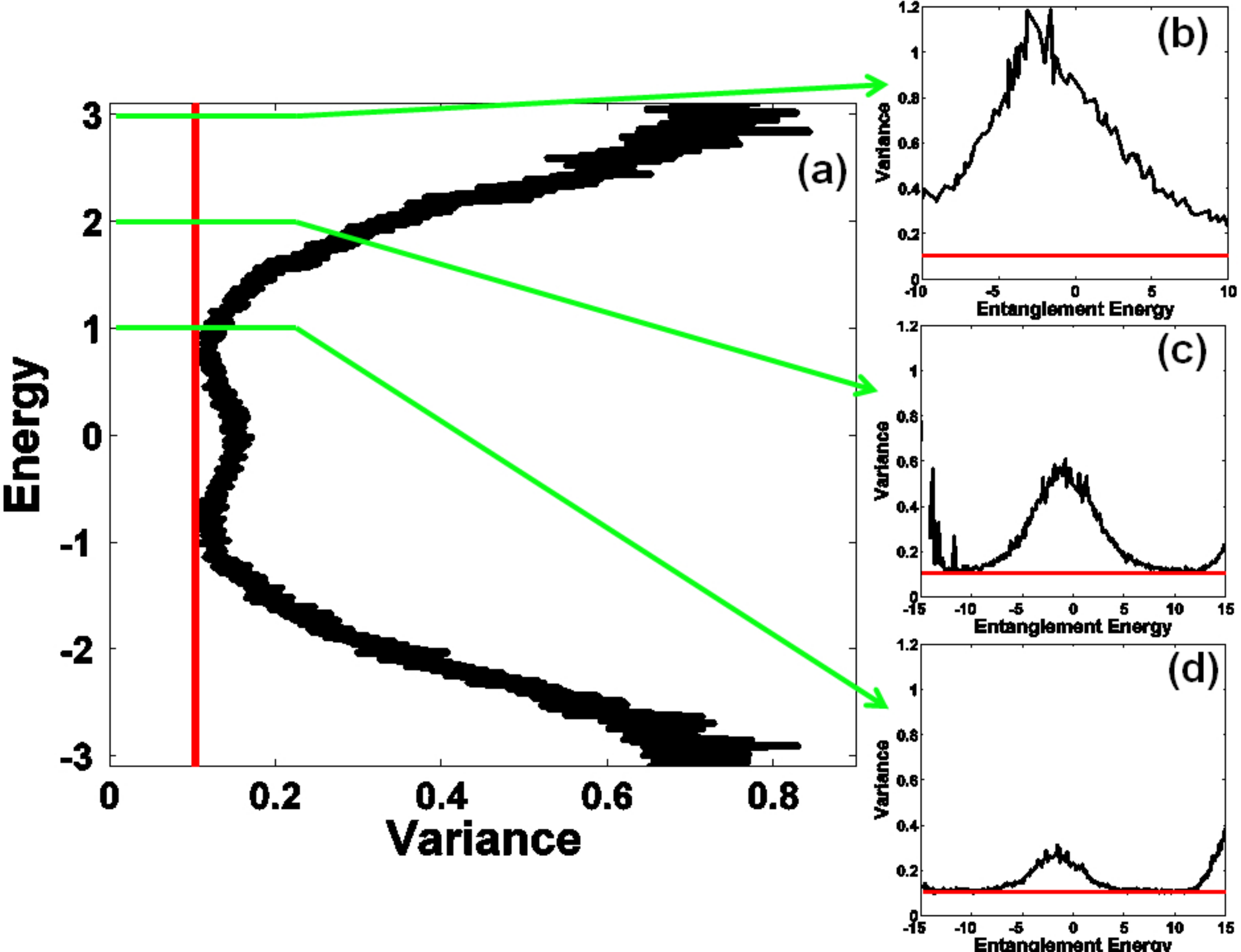}\caption{\label{fig:varslice}(a) Plot of the level statistics of a QSH insulator with a BIA term of $\lambda=0.3$ and a random disorder potential of magnitude $W=5$. In plots (b)-(d) we show the resultant entanglement spectrum calculated at chemical potentials of (b) $\mu=3$ (c) $\mu=2$ (d) $\mu=1$}
\end{figure}
\begin{figure}[t]
\centering{}\includegraphics[width=0.5\textwidth]{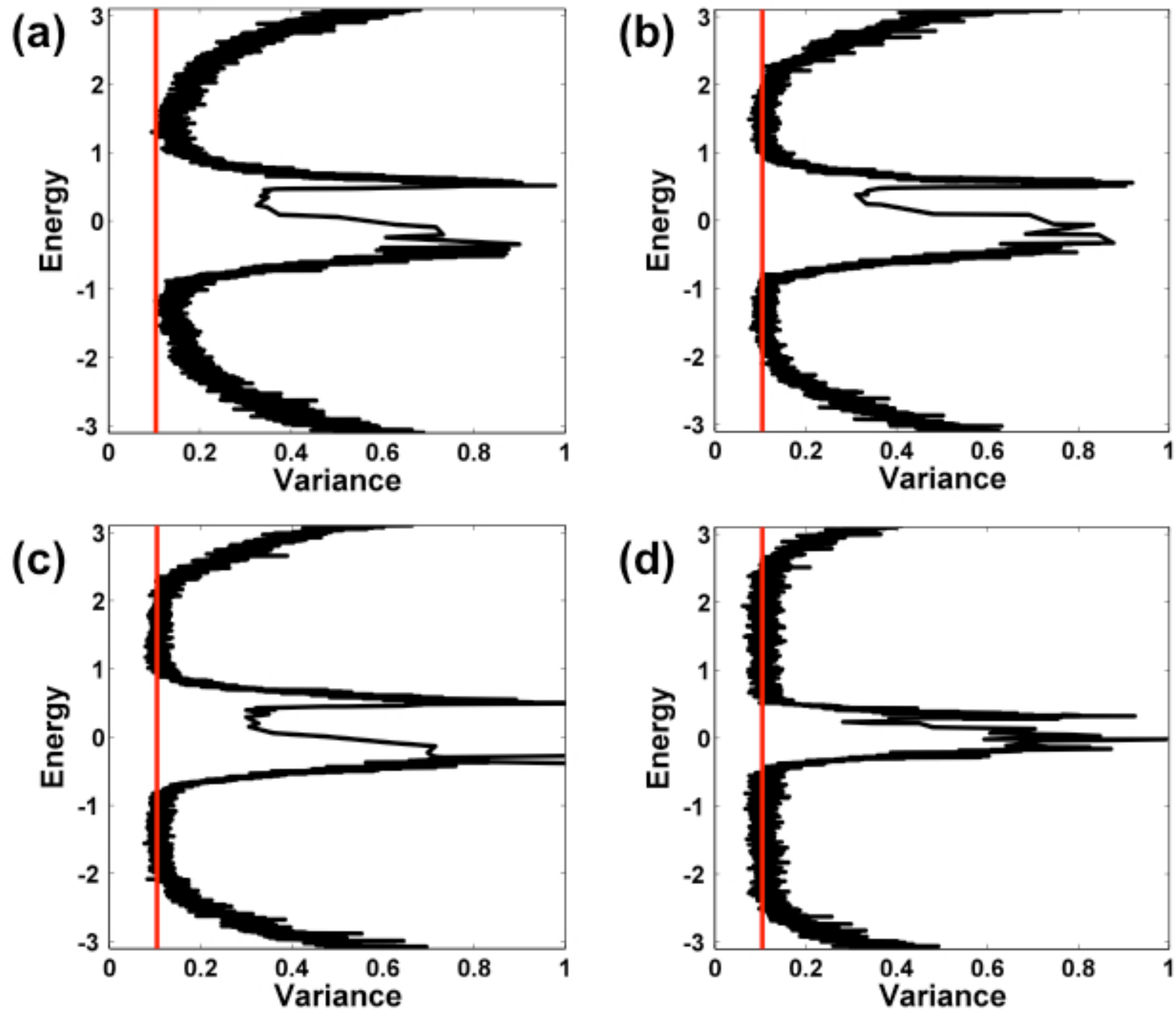}\caption{\label{fig:biaslice}Plot of the level statistics of a QSH insulator with $W=1$ for various values of the BIA term. (a)$\lambda=0.0$ (b)$\lambda=0.0$ (c)$\lambda=0.0$ (d)$\lambda=0.0$ We clearly see that as $\lambda$ increases the width of the delocalized region of states increases proportionally.}
\end{figure}
Let us first examine the nearest-neighbor level-spacing statistics for the energy spectrum\cite{mehtabook}. To calculate the distribution of level spacings one picks an energy $E_0$ in the spectrum and calculates the nearest-neighbor level spacings of states in a small window around $E_0$ and then combines the level spacing data for many different random disorder configurations to form a distribution for each choice of $E_0$. In this work, we use a window which considers states $3$ levels above and below $E_0$. In Fig. \ref{fig:varslice}(a) we plot the normalized variance  $(\langle s^{2} \rangle - \langle s \rangle^{2})$ of this distribution for a large range of choices for $E_0$ covering the entire spectrum of the Hamiltonian. This data was taken for $W=5$ and $\lambda =0.3$ and averaged over 1000 disorder configurations. Our model is particle-hole symmetric (but only on average disorder is included) and the resultant distribution variances are also symmetric. For class AII we expect two different types of signatures: (i) a Poisson distribution ($P_{P}=e^{-s}$) for states which are localized with a variance which approaches $1$ in finite-size (ii) a symplectic Wigner-Surmise distribution ($P_{GSE}=Ns^4 \exp \left(-\frac{64s^2}{9\pi}\right)$, with $N$ a normalization) with a variance of which approaches $\sim 0.1045.$ We see evidence for both types of distributions depending on the value of the chemical potential. Near the band edges, we see the variance trending toward $1$; in fact, if the system size is increased the variance away from the center of the spectrum more quickly asymptotes to a value of $1$ indicating mostly localized states. However, there are two regions (near $E_0=\pm 1$) which have variances very close to the Wigner-Surmise value (red-vertical line). Compared to similar data of two of the authors shown in Ref. \onlinecite{prodan2010} we notice a difference in that there seems to be a relatively wide band of states that reach variance $\sim 0.1045$ compared to a much narrower range for the Chern insulator. We take this to be a signature of a critical metallic region (as opposed to a critical point) separating the trivial from topological insulator phases. For the disordered AII class such a region is expected\cite{prodan2011}. As an aside let us note that if we tune $\lambda\to0$ the delocalized region thins, and if we increase $\lambda$ (as shown in Fig. \ref{fig:biaslice} for $W=1$) the width increases which shows that the band of delocalized states is due to the symplectic nature of the random matrices. Thus for class AII one must pass through an intermediate metallic/critical phase when going from the disordered topological insulator phase to the disordered trivial insulator phase via the levitation and annihilation transition mechanism.

Now this behavior is to be compared with the resultant entanglement spectra of the same system with ground-state fillings given by several chemical potentials shown in Figs \ref{fig:varslice}(b)-(d). Here we see that the EnS also has a regions with large variances of order $1$ and regions with variances that closely approach $0.1045.$  In Fig.~\ref{fig:varslice}(b) the filling at $\mu=3$ is such that the states at the Fermi-level are localized, and the system is in a trivial insulator phase. We see that the variance of the entanglement energy levels (denoted $\xi_i$) derived from the eigenvalues of the truncated $C_{ij}$ (denoted $\zeta_i$) result in a large variance and exhibit a Poisson-like distribution for essentially every choice of the entanglement energy $\xi_0.$ At $\mu=2$ the picture is less clear but it appears that some regions of entanglement energy are approaching the delocalized distribution, though the curves do not actually touch the asymptotic value. This does not just seem to be a finite-size effect since as we approach closer to the delocalized band of states the variances actually get much closer to the asymptotic Wigner-Surmise value. For $\mu=1$ where the filling is very nearly in the band of delocalized states we see the variance of the entanglement energy spacings flattening out to the delocalized value. It should be noted that a similar effect was seen in the case of the Chern insulator where, when tuned to into the band of delocalized states the variance of the entanglement energies completely flattened out to the Wigner-surmise value\cite{prodan2011}. It appears that this phenomenon is a general property (at least of topological insulators) as we tune the chemical potential through delocalized states. We have verified that this behavior with larger system sizes of $50\times 50$ with a corresponding $25\times 25$ region A in order to rule out the possibility of significant finite size effects skewing the results of the entanglement.
We also note that all of the entanglement variances have a similar shape with a peak near $\xi=0.$ This property also occurred for the Chern insulator and is not completely understood. Our intuition indicates that this peak is arising from ``local entanglement." States that give rise to the low energy part of the entanglement spectrum  (\emph{i.e.} those states which are highly entangled) arise from two mechanisms: (i) they are delocalized states  which have significant probability to be in region A and its compliment (ii) they are localized states which, by chance, are localized near the cut and contribute to the entanglement in spite of their local nature. A method to remove local entanglement from these calculations is still lacking, but in principle should make these calculations easier to interpret.

\begin{figure}[t]
\centering{}\includegraphics[width=0.48\textwidth]{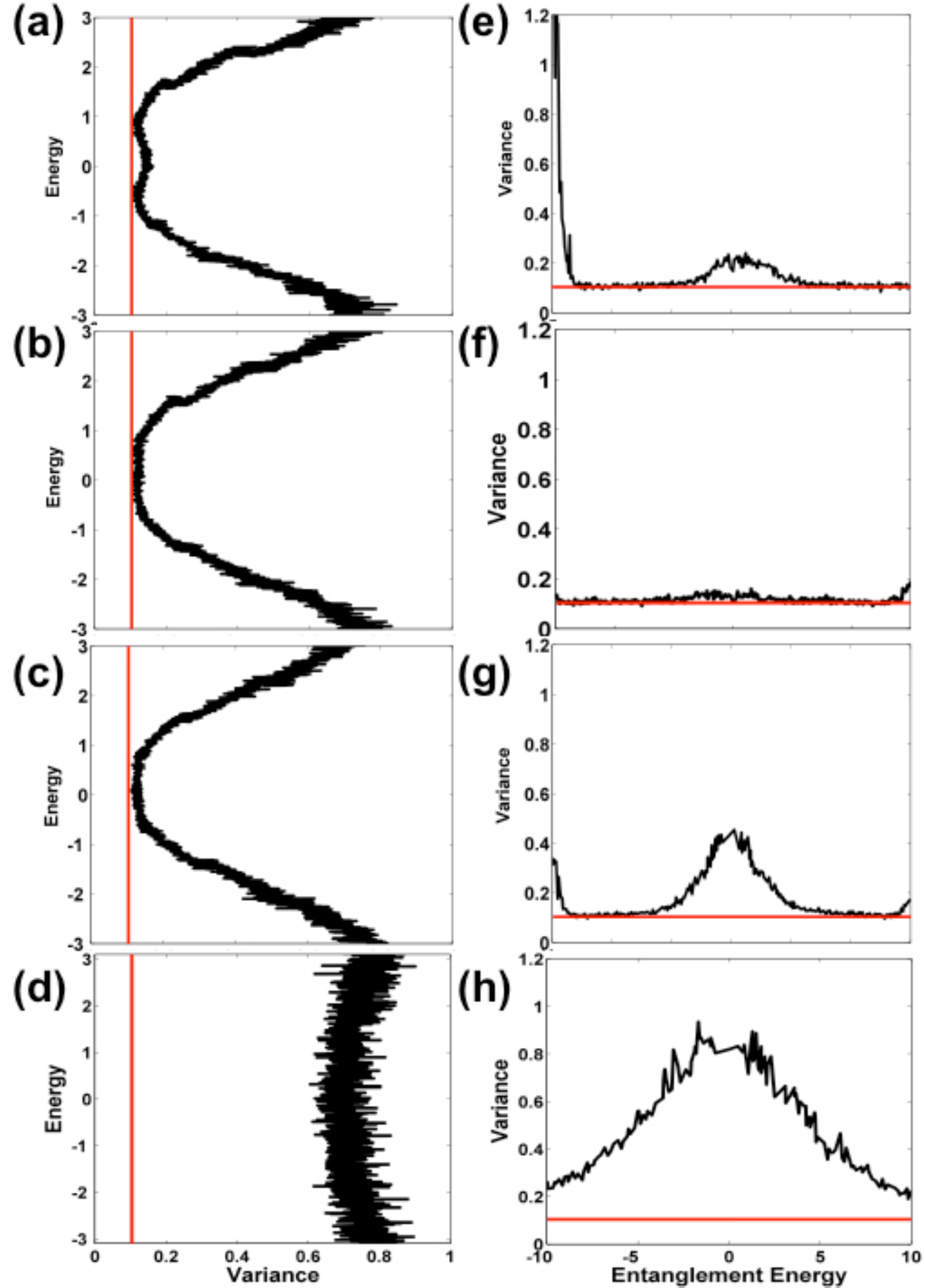}
\caption{\label{fig:mvar}Plots of the variance of the level spacings for the full spectrum corresponding to random disorder strengths of (a) $W$=5.1, (b) $W$=5.625, (c) $W$=6 and (d) $W$=10. We see that as disorder strength is increased we pass the quantum critical point and the system passes from topological to metallic then to insulating. In (e)-(h), we plot the entanglement spectra for $\mu$=0 corresponding to the disorder strengths considered in (a)-(d) with a BIA term of $\lambda$=0.3.}
\end{figure}
Having considered the properties of the entanglement statistics as a function of chemical potential we now wish to examine the dual case of a fixed chemical potential but varying disorder strength (still at $\lambda=0.3$). To calculate the entanglement spectrum we will fix our ground-state filling at $\mu=0$ and focus near the region of disorder strength where the system exhibits phase transitions between a topological insulator, metal, and trivial insulator at this energy. For our model this occurs when the disorder strength $W$ satisfies $5<W<6.$ We show energy and corresponding entanglement variances (for a filling of $\mu=0$) in Fig. \ref{fig:mvar}. For Figs. \ref{fig:mvar}(a), (b), (c) and (d), we see the variances for the energy spectrum in the topological insulator ($W=5.1$), metal ($W=5.625$), and trivial insulator ($W=6$ and $W$=10) phases respectively. We estimate the critical point separating the metal phase from the trivial insulator occurs around $W=5.85$ which is the point where the energy variance curve near $E=0$ lifts off the delocalized value. The phase transition occurs as the two bands of delocalized states merge near $E=0$ and eventually annihilate in a manner consistent with the scenario of levitation and pair annihilation\cite{Onoda2007}. As these are clearly \emph{bands} of delocalized states there is a finite range of $W$ over which these states are annihilating which gives rise to the critical metallic region. We see clear features in the corresponding entanglement spectra, with the most interesting one seen in Fig. \ref{fig:mvar}(f) which shows a flat variance curve for the entanglement energies in the case where the chemical potential intersects the critical metallic region at $E=0.$ The variances for the two insulating phases in Figs. \ref{fig:mvar}(e) (topological) and (g) (trivial) may be distinguished by noting that the topological phase contains regions of extended states touching the prescribed variance while the trivial phase exhibits bands of states which have a variance which sits above the requisite variance. This is exacerbated in Fig. \ref{fig:mvar}(h) where the entanglement spectrum sits well above the variance denoting the presence of delocalized states. Figs. \ref{fig:mvar}(d) and (f) show evidence of contributions from local entanglement which produces the central maximum. This maximum seems to be a feature of the gapped phases as it does not appear when the system is tuned to a critical metallic region either by varying chemical potential, or in this case, varying disorder strength until a critical region hits the fixed $\mu.$ 

\begin{figure}[t]
\centering{}\includegraphics[width=0.48\textwidth]{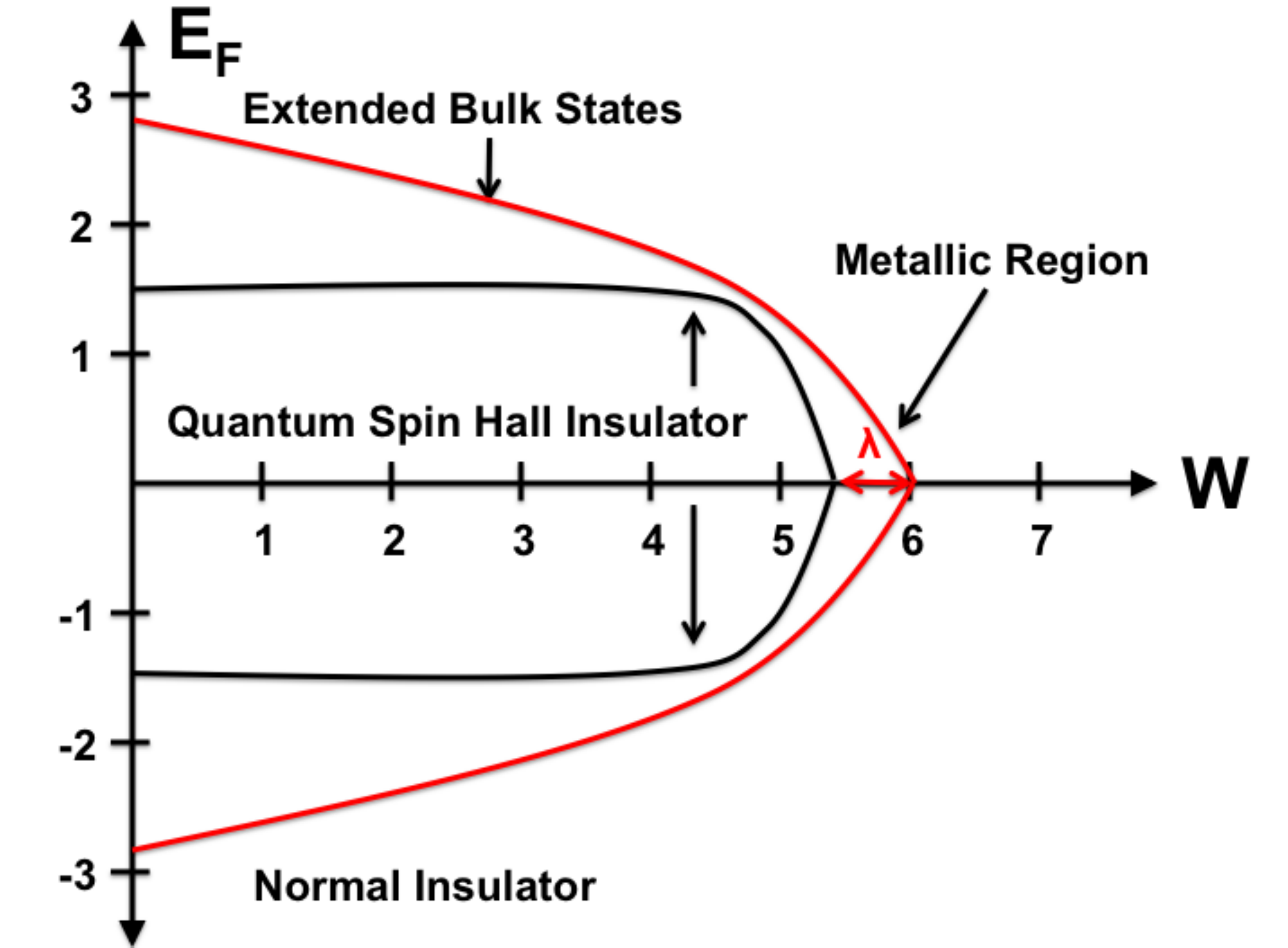}\caption{\label{fig:phasediag}Phase diagram of the disordered AII class as a function of disorder strength $W$ and Fermi level $E_F$ for $\lambda=0.3.$ There is a connected region of quantum spin Hall insulator separated from the outer trivial insulator phase by a metallic critical region of delocalized states. The width of this region in energy is proportional to $\lambda$ (the BIA strength) and is not constant as $W$ is varied.}
\end{figure}
As hinted in Ref. \cite{prodan2010}, and shown to be a more general feature in this work, the flattening of the variance of the entanglement energy spacing is an indicator that the filling level is tuned to a region of delocalized states and can be used to identify critical regions effectively in order to map out a phase diagram.  
Using this technique, we can glean information on the resultant topological phase transitions from a careful analysis of the EnS. We are now in a position to discuss the phase diagram of our disordered QSH model.  In Fig.~\ref{fig:phasediag}, we plot the phase diagram as a function of random disorder strength and chemical potential for $\lambda = 0.3$. We see a connected region of the quantum spin Hall phase separated from an outer region of trivial insulator by an intervening metallic region. The width of the metallic region in energy  appears to decrease somewhat over a range of increasing disorder strengths.  When the BIA term is increased the metallic region persists over a similar increase in disorder strength, and the width of the metallic region in energy also increases by a similar amount. 

To summarize, we have shown that the variance of the nearest-neighbor level spacing distribution of the entanglement spectrum can serve as a good indicator for regions of energy which contain delocalized states. We have illustrated the nature of the symplectic AII class by showing the existence of wide bands of extended states with an energy width which depends on the strength of the inversion breaking perturbation. While issues with local entanglement due to localized states likely obscure some features of the entanglement spectra, a generic feature of a flat level-spacing variance is indicative of the Fermi level being tuned to a critical region of extended states. This observation will be helpful in identifying localized/de-localized phase boundaries in disordered fermion systems. 

\begin{acknowledgments}
We acknowledge useful conversations with E. Prodan, G. Fiete, M. B. Hastings, R. Thomale and B. Dellabetta. MJG acknowledges support from the AFOSR under grant FA9550-10-1-0459 and the ONR under grant N0014-11-1-0728. BAB was supported by NSF CAREER DMR- 095242, ONR - N00014-11-1-0635, Darpa - N66001-11- 1-4110, the Keck Foundation and David and Lucile Packard Foundation. TLH acknowledges support from the NSF under grant DMR 0758462 at the University of Illinois.
\end{acknowledgments}
\bibliography{TI}

\end{document}